\documentclass[apj,numberedappendix]{emulateapj}
\usepackage{apjfonts}
\usepackage{graphicx,url,amssymb,amsmath,longtable,rotating,color,units,
wasysym,subfigure,epsfig,listings,bm}

\newcommand{\be}{\begin{eqnarray}}
\newcommand{\ee}{\end{eqnarray}}

\newcommand{\lp}{\left(}
\newcommand{\rp}{\right)}
\newcommand{\lb}{\left[}
\newcommand{\rb}{\right]}

\begin{document}

\slugcomment{Accepted for publication in The Astrophysical Journal Letters}
\shorttitle{Taking the ``Un'' out of ``Unnovae''}
\shortauthors{Piro, A. L.}

\normalsize


\title{Taking the ``Un'' out of ``Unnovae''}

\author{Anthony L. Piro}

\affil{Theoretical Astrophysics, California Institute of Technology, 1200 E California Blvd., M/C 350-17, Pasadena, CA 91125; piro@caltech.edu}


\begin{abstract}
It has long been expected that some massive stars produce stellar mass black holes (BHs) upon death. Unfortunately, the observational signature of such events has been unclear. It has even been suggested that the result may be an ``unnova,'' in which the formation of a BH is marked by the disappearance of a star rather than an electromagnetic outburst. I argue that when the progenitor is a red supergiant, evidence for BH creation may instead be a $\approx3-10\,{\rm day}$ optical transient with a peak luminosity of $\approx10^{40}-10^{41}\,{\rm erg\,s^{-1}}$, a temperature of $\approx10^4\,{\rm K}$, slow ejection speeds of $\approx200\,{\rm km\,s^{-1}}$, and a spectrum devoid of the nucleosynthetic products associated with explosive burning. This signal is the breakout of a shock generated by the hydrodynamic response of a massive stellar envelope when the protoneutron star loses $\sim\ {\rm few}\times0.1\,M_\odot$ to neutrino emission prior to collapse to a BH. Current and future wide-field, high-cadence optical surveys make this an ideal time to discover and study these events. Motivated by the unique parameter space probed by this scenario, I discuss more broadly the range of properties expected for shock breakout flashes, with emphasis on progenitors with large radii and/or small shock energies. This may have application in a wider diversity of explosive events, from pair instability supernovae to newly discovered but yet to be understood transients.
\end{abstract}

\keywords{black hole physics ---
	stars: evolution  ---
	stars: transients ---	
	supernovae: general}


\section{Introduction}
\label{sec:introduction}

It is currently unknown what fraction of massive stars produce black holes (BHs) rather than neutron stars (NSs), what the channels for BH formation are, and what corresponding observational signatures are expected. There is strong evidence for stellar mass BHs from X-ray binaries throughout our galaxy \citep{RemillardMcClintock2006}, so it is clear BHs must be a possible endpoint of stellar evolution. Pre-explosive imaging of core-collapse supernovae (SNe) suggests progenitor masses $\lesssim17-20\,M_\odot$ \citep{Smarttetal2009} for standard Type II-P SNe, which are thought to produce NSs. Assuming a Salpeter initial mass function, this implies that an upper limit of $\sim30-35\%$ of massive stars above $8\,M_\odot$ fail to lead to a successful SNe, and perhaps this number is related to the fraction of BHs produced. Such a straightforward comparison is complicated by the impact of binary interactions \citep{Smithetal2011}, which are expected to dominate the evolution of most massive stars \citep{Sanaetal2012}. A collapsar and gamma-ray burst likely accompanies some instances of BH formation \citep[e.g.,][]{MacFadyenWoosley1999}, but these are too rare to explain most BHs and are confined to certain environments \citep{Staneketal2006,Modjazetal2008}. It is possible that the signature of BH formation is in fact the disappearance of a massive star, or ``unnova,'' rather than an actual SN-like event \citep{Kochaneketal2008}.

On the theoretical side there is also much uncertainty. Work by \citet{Timmesetal1996}, \citet{Fryer1999}, \citet{Hegeretal2003}, \citet{EldridgeTout2004}, \citet{Zhangetal2008}, \citet{OConnorOtt2011}, and \citet{Uglianoetal2012} attempts to connect the outcomes of stellar collapse to the progenitor zero-age main sequence (ZAMS) mass and metallicity. At solar metallicities, such models predict BH formation for roughly $\sim10-25\%$ of progenitors, but this depends sensitively on many uncertain factors such as the treatment of mass loss. Neutrino emission may be a way of inferring collapse to a BH \citep{Burrows1986,Baumgarteetal1996,Liebendorferetal2004,Sumiyoshietal2007}, but this will only be detected for especially nearby events. Rotation may assist in producing a SN-like signature \citep{WoosleyHeger2012,DexterKasen2012} and this may enhance gravitational wave emission \citep{PiroThrane2012}, but it is not clear in what fraction of events rotation is important. If indeed the collapse proceeds as a simple implosion, then the star is not expected to brighten significantly \citep{Shapiro1989,Shapiro1996}, and an unnova-like signature is again implied.

It is important to remember that massive stars which hydrostatically form degenerate iron cores never directly collapse to BHs \citep[e.g.,][]{Burrows1988,OConnorOtt2011}. BH formation is always preceded by a protoneutron star phase with abundant emission of neutrinos \citep{Burrows1988,Beacometal2001} and gravitational waves \citep{Ott2009} until the protoneutron star  contracts within its event horizon. In a somewhat forgotten theoretical study, \citet{Nadezhin1980} focused on the impact of the loss of $\sim{\rm few}\times0.1M_\odot$ over several seconds from this neutrino emission. The star reacts as if the gravitational potential of the core has abruptly changed and expands in response. This develops into a shock propagating into the dying star's envelope. Although the shock's energy of $E\sim10^{47}-10^{48}\,{\rm erg}$ is relatively small in comparison to typical core-collapse SNe, it can nevertheless eject the loosely bound envelope of a red supergiant. This idea was recently investigated in more detail by \citet{LovegroveWoosley2013}. Their study focused on the SN-like transient that results. Their general conclusion was that such an event would be dominated by recombination of hydrogen, similar to Type II-P SNe. This would produce a plateau-like light curve lasting a year or more with a luminosity of $\sim~{\rm few}\times10^{39}\,{\rm erg\,s^{-1}}$ and a cool temperature of $\lesssim4000\,{\rm K}$. So although this may be a more promising counterpart to BH formation than an unnova, it would still be challenging to identify with current observational capabilities.

Motivated by these previous studies, I investigate in more detail the shock breakout expected from this mechanism. This demonstrates that the breakout emission may be the most promising signature of BH formation from a red supergiant. In \S \ref{sec:estimates}, I estimate the main properties of the shock breakout, and discuss more broadly the emission properties at large radii and/or low energies. I conclude in \S \ref{sec:conclusion} with a summary of my results and a discussion of future work.

\section{Shock Breakout Estimates}
\label{sec:estimates}

I begin by briefly reviewing the physics of shock breakout. For a more detailed background, the interested reader should refer to the abundant literature on this topic (including, but not limited to \citealp{ImshennikNadezhin1989}; \citealp[][hereafter MM99]{MatznerMcKee1999}; \citealp{NakarSari2010,NakarSari2012,Katzetal2012}). The main difference in this present work is the relatively low energy of the shock.

The envelope is approximated as a polytrope with index $n$ given by $\rho_0 = \rho_1( R_*/r-1)^n$, where $\rho_1$ is the half-radius density, and I focus on $n=3/2$, as is appropriate for a convective envelope. As is common practice, I scale $\rho_1$ by the characteristic density of the ejecta $\rho_*=M_{\rm ej}/R_*^3$, where $M_{\rm ej}$ is the mass of the ejecta, since the ratio $\rho_1/\rho_*$ is typically of order unity\footnote{See the Appendix of \citealp{CalzavaraMatzner2004} for derivations of values for $\rho_1/\rho_*$ for different stellar structures.}. For the very outer parts of the star, I use the variable $x=1-r/R_*$ and set $\rho_0 \approx \rho_1 x^n$, where $x\ll1$. The shock propagates through the envelope with speed $v_s = \Gamma (E/m)^{1/2}(m/\rho_0 r^3)^\beta$ (MM99), where $E$ is the shock energy, $r$ is the radial coordinate, $m$ is the mass interior to $r$, and $\rho_0$ is the density profile of the envelope prior to expansion due to the shock. Using the self-similar, planar solutions of \citet{GandelManFrank1956} and \citet{Sakurai1960}, I adopt the values of $\beta=0.19$ and $0.23$ when the shock is radiation-pressure and  gas-pressure dominated, respectively. The constant $\Gamma$ can be estimated from self-similar blastwave solutions (MM99), for which I use $\Gamma=0.794$.

Breakout happens when a shock gets too close to the stellar surface, and photons diffuse out into space at an optical depth $\tau \approx c/v_s$. Setting $\tau = \kappa \rho_1R_*x^{n+1}/(n+1)$, breakout occurs at
\be
	\tau_{\rm bo} = \frac{c}{\Gamma v_*}
	\lb
		\lp\frac{\rho_1}{\rho_*}\rp^{-1/n}
		\frac{\Gamma}{n+1}
		\frac{v_*}{c}
		\frac{\kappa M_{\rm ej}}{R_*^2}
	\rb^{-\beta/(1+1/n-\beta)},
\ee
where $v_*=(E/M_{\rm ej})^{1/2}$. Scaling this to typical values for a low energy shock in a massive progenitor,
\be
	\tau_{\rm bo} = 2.9\times10^3
	\frac{M_{10}^{0.44}R_{1000}^{0.26}}{\kappa_{0.34}^{0.13}E_{48}^{0.56}}
	\lp\frac{\rho_1}{\rho_*} \rp^{0.086},
\ee
where $\kappa_{0.34}=\kappa/0.34\,{\rm cm^2\,g^{-1}}$, $E_{48}=E/10^{48}\,{\rm erg}$, $M_{10}=M_{\rm ej}/10\,M_\odot$, and $R_{1000}=R_*/1000\,R_\odot$.

The amount of energy in the radiation field associated with the shock depends on whether or not the shock is dominated by radiation pressure. For the radiation-dominated case, the shock jump condition is $aT^4/3=6\rho_0v_s^2/7$, where $a$ is the radiation constant. Setting the energy to be $E_{\rm rad}\approx 4\pi R_*^3x_{\rm bo}(aT^4/3)$, where $x_{\rm bo}$ is the depth at which $\tau=\tau_{\rm bo}$,
\be
	E_{\rm rad} = 1.2\times10^{47}
	\frac{E_{48}^{0.56}R_{1000}^{1.74}}{\kappa_{0.34}^{0.87}M_{10}^{0.44}}
	\lp\frac{\rho_1}{\rho_*} \rp^{-0.086}{\rm erg}.
\ee
If the shock is gas-dominated then $\rho_0k_{\rm B}T/\mu m_p=3\rho_0 v_s^2/16$, where $k_{\rm B}$ is Boltzmann's constant.
In this case\footnote{Note that although I denote this energy with a subscript ``gas,'' this energy is still associated with the radiation field. This merely identifies that gas pressure is dominant in this regime.}
\be
	E_{\rm gas} = 1.7\times10^{48}
	\frac{E_{48}^{4.4}R_{1000}^{1.37}}{\kappa_{0.34}^{1.74}M_{10}^{3.59}}
	\lp \frac{\rho_1}{\rho_*}\rp^{-1.20}{\rm erg}.
\ee
The actually shock breakout energy is roughly given by $E_{\rm bo}\approx\min(E_{\rm rad},E_{\rm gas})$. A key point is the strong scalings of $E_{\rm gas}\propto E^{4.4}$, which suppresses shock breakout for small $E$. For this reason, all equations in this paper with numerical factors assume the radiation-dominated regime unless otherwise noted.

The observed luminosity is determined by the timescale over which this energy is emitted. The two dominant effects are the light-travel time \citep{EnsmanBurrows1992},
\be
	t_{\rm lt} \approx R_*/c = 39\,R_{1000}\,{\rm min},
\ee
and the diffusion time (MM99; Piro \& Nakar 2012),
\be
	t_{\rm diff} = R_*x_{\rm bo}\tau_{\rm bo}/c
	= 9.6
	\frac{M_{10}^{0.21}R_{1000}^{2.16}}{\kappa_{0.34}^{0.58}E_{48}^{0.79}}
	\lp\frac{\rho_1}{\rho_*} \rp^{-0.28}{\rm days}.
	\label{eq:tdiff}
\ee
The timescale for the breakout emission is $t_{\rm bo}\approx \max(t_{\rm lt},t_{\rm diff})$. Note that $t_{\rm diff}>t_{\rm lt}$ when
\be
	E<1.7\times10^{51}
	\frac{M_{10}^{0.27}R_{1000}^{1.47}}{\kappa_{0.34}^{0.73}}
	\lp\frac{\rho_1}{\rho_*} \rp^{-0.35}{\rm erg},
\ee
so that for the low energy shock investigated here $t_{\rm bo}\approx t_{\rm diff}$. If $t_{\rm diff}$ is too long, then the breakout is suppressed by cooling from adiabatic expansion. This occurs on a timescale,
\be
	t_{\rm exp} = R_*/v_f= 38\frac{M_{10}^{0.44}R_{1000}^{1.26}}{\kappa_{0.34}^{0.13}E_{48}^{0.57}}
	\lp\frac{\rho_1}{\rho_*} \rp^{0.086}{\rm days},
\ee
where $v_f$ is the final velocity of the material
\be
	v_f = 210
	\frac{\kappa_{0.34}^{0.13}E_{48}^{0.57}}{M_{10}^{0.44}R_{1000}^{0.26}}
	\lp\frac{\rho_1}{\rho_*} \rp^{-0.086}
	{\rm km\,s^{-1}},
	\label{eq:vmax}
\ee
which includes a factor of 2 from geometric effects and pressure gradients (MM99).

An additional issue is thermalization \citep[see the discussion in][]{NakarSari2010}. When thermal emission is achieved at temperature $T$, the number density of photons is $n_{\rm ph}\approx a T^3/3k_{\rm B}$. The rate of photon production from free-free emission is $\dot{n}_{\rm ph}\approx3.5\times10^{36}\rho^2T^{-1/2}\,{\rm s^{-1}\,cm^{-3}}$. Therefore thermalization is expected for times later than roughly
\be
	t_{\rm therm} \approx \frac{n_{\rm ph}}{\dot{n}_{\rm ph}}
	\approx 0.03
	\frac{	\kappa_{0.34}^{1.01}E_{48}^{1.37}R_{1000}^{1.39}}{M_{10}^{1.51}}
	\lp\frac{\rho_1}{\rho_*} \rp^{1.19}{\rm s}.
\ee
Since $t_{\rm therm}\ll t_{\rm lt}, t_{\rm diff}$ at low energies, the shock breakout is in the thermal regime. This is in contrast to typical SNe with an energy of $\sim 10^{51}\,{\rm erg}$ where $t_{\rm bo}\approx t_{\rm lt}$ and $t_{\rm therm}\gtrsim t_{\rm bo}$.

\begin{figure}
\epsscale{1.18}
\plotone{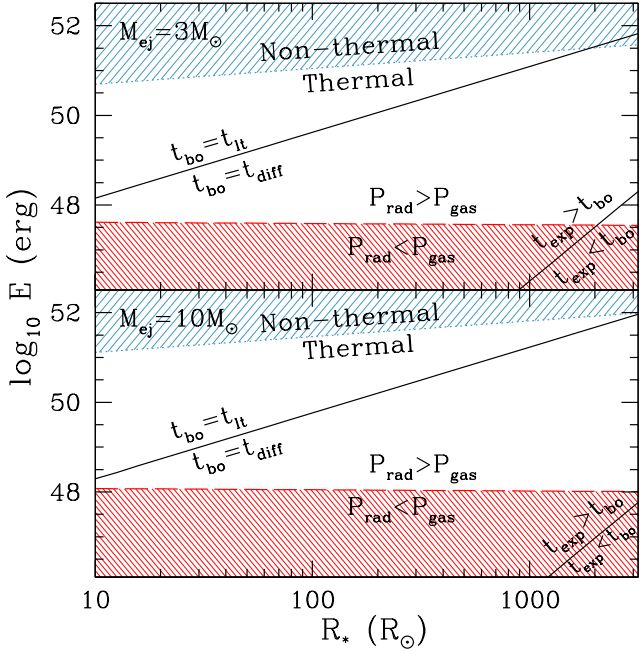}
\caption{Main physical conditions that determine shock breakout as a function of $R_*$ and $E$. Top and bottom panels are ejecta masses of $3\,M_\odot$ and $10\,M_\odot$, respectively. The solid line divides regions where the duration $t_{\rm bo}$ is determined by either the light-crossing time $t_{\rm lt}$ or the thermal diffusion time $t_{\rm diff}$. Above the dotted line is the non-thermal regime (blue, lightly shaded). Below the dashed line the shock becomes gas-pressure dominated (red, darkly shaded). In the far right bottom corner, $t_{\rm exp}<t_{\rm bo}$ and the breakout is suppressed by adiabatic expansion.}
\label{fig:energyradius}
\epsscale{1.0}
\end{figure}

In Figure \ref{fig:energyradius}, I highlight the main physical processes that determine the shock breakout emission as a function of $R_*$ and $E$. The upper panel is for an ejecta mass of $M_{\rm ej}=3\,M_\odot$ and the lower panel is for $M_{\rm ej}=10\,M_\odot$. The diagonal solid line divides where the breakout timescale is determined by $t_{\rm lt}$ or $t_{\rm diff}$. The dotted line shows where the emission is expected to transition to being non-thermal (blue, lightly shaded). The dashed line shows where the shock starts to be gas pressure dominated (red, darkly shaded).

\begin{figure}
\epsscale{1.18}
\plotone{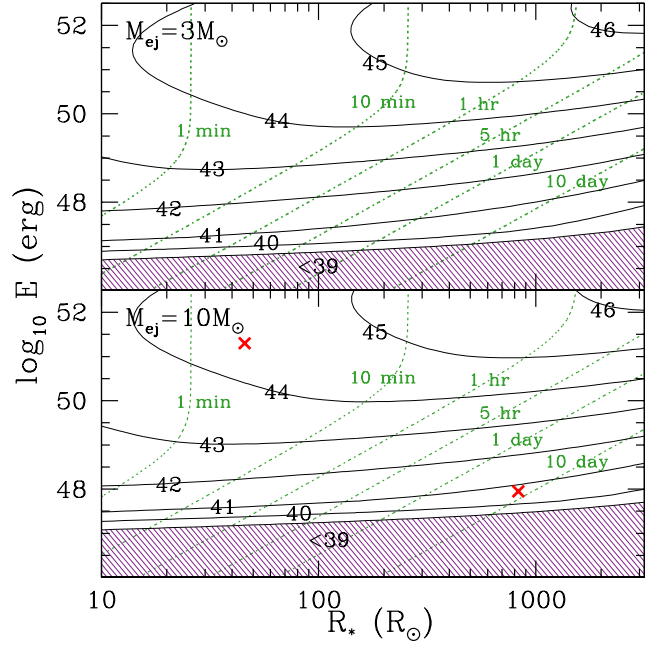}
\caption{Same as Figure \ref{fig:energyradius}, this time plotting contours of constant breakout luminosity $L_{\rm bo}$ (solid lines) and constant duration $t_{\rm bo}$ (green, dotted lines). The contours are labeled with their values, with the contours of constant luminosity labeled by $x$ where $L_{\rm bo}=10^x\,{\rm erg\,s^{-1}}$. Purple, shaded regions denote where $L_{\rm bo}<10^{39}\,{\rm erg\,s^{-1}}$. The decrease in $L_{\rm bo}$ at especially large $R_*$ and $E$ (the upper-left and lower-right corners) is due to adiabatic expansion. The upper red cross shows a characteristic model from \citet{EnsmanBurrows1992}, which is appropriate for core-collapse of a blue supergiant. The lower red cross corresponds to BH formation of a $15\,M_\odot$ ZAMS red supergiant \citep[][maximum mass loss model]{LovegroveWoosley2013}.}
\label{fig:luminosity}
\epsscale{1.0}
\end{figure}

In Figure \ref{fig:luminosity}, I show how these different physical conditions translate into observed luminosities and durations. Solid lines denote constant breakout luminosity, which is set as
\be
	L_{\rm bo} = \min(E_{\rm rad},E_{\rm gas})/\max(t_{\rm lt},t_{\rm diff}),
\ee
where I add an interpolation in each of the $\min$ and $\max$ functions to smooth the transitions. Furthermore, I include a factor of $[1+(t_{\rm bo}/t_{\rm exp})^3]^{-\gamma}$ to account for adiabatic expansion \citep{Piroetal2010}, where $\gamma$ is the adiabatic exponent. The solid lines are labeled by the $x$, where $L_{\rm bo}=10^x\,{\rm erg\,s^{-1}}$. The breakout luminosity decreases rapidly once gas pressure dominates, and I shade regions where $L_{\rm bo}<10^{39}\,{\rm erg\,s^{-1}}$. Dotted lines donate constant breakout durations $t_{\rm bo}=\max(t_{\rm lt},t_{\rm diff})$. Red crosses compare core-collapse of a blue supergiant \citep{EnsmanBurrows1992} with BH formation from a red supergiant, demonstrating the different regimes these two cases occupy.

Since it may be generally useful, a wider range of $R_*$ and $E$ are plotted in Figures \ref{fig:energyradius} and \ref{fig:luminosity} than just what applies to the BH formation case that is the focus here. For example, for a pair instability SN with $R_*\sim2000\,R_\odot$ and $E\sim10^{52}-10^{53}\,{\rm erg}$, the expectation from Figure \ref{fig:luminosity} is that $L_{\rm bo}\sim10^{46}\,{\rm erg\,s^{-1}}$ with $t_{\rm bo}\sim{\rm hrs}$. This is roughly consistent with more detailed calculations\footnote{The progenitors of pair instability SNe are considerably more massive, but this is easily corrected for by using the scalings provided in this work.} \citep{Kasenetal2011}. Similarly, if the luminosity and duration of a purported breakout flash is observed, these results can be used to infer $R_*$ and $E$, potentially even providing constraints on transient events that are difficult to classify.

Especially relevant to the situation of BH formation is the regime where $E_{\rm bo}\approx E_{\rm rad}$ and $t_{\rm bo}\approx t_{\rm diff}$, which results in
\be
	L_{\rm bo} = 1.4\times10^{41}
	\frac{E_{48}^{1.36}}{\kappa_{0.34}^{0.29}M_{10}^{0.65}R_{1000}^{0.42}}
	\lp\frac{\rho_1}{\rho_*} \rp^{0.194}{\rm erg\,s^{-1}}.
	\label{eq:lbo}
\ee
This applies to the entire lower-right triangle of parameter space in each of the panels of Figure \ref{fig:energyradius} (although note that adiabatic expansion can cause the luminosity to be somewhat lower than this).
The observed shock breakout temperature is in the thermal regime and thus can be estimated as $T_{\rm obs}\approx T_{\rm bo}/\tau_{\rm bo}^{1/4}$, where $T_{\rm bo}$ is the temperature of the plasma at the depth of the breakout, resulting in
\be
	T_{\rm obs}=1.4\times10^4
	\frac{E_{48}^{0.34}}{\kappa_{0.34}^{0.068}M_{10}^{0.16}R_{1000}^{0.61}}
	\lp\frac{\rho_1}{\rho_*} \rp^{0.049}
	{\rm K}.
\ee
Since $T_{\rm bo}\gtrsim T_{\rm obs}$, an electron scattering opacity is sufficient to estimate the diffusion properties. At the surface, where the temperature is close to $T_{\rm obs}$, other opacity effects may impact the observed spectra, which should be modeled in the future.

Subsequent to the breakout emission, there is a plateau-like phase due to hydrogen recombination as investigated by \citet{LovegroveWoosley2013}. Using the analytic fits to the numerical models of Type II-P SNe by \citet{KasenWoosley2009}, I estimate that this phase has a luminosity
\be
	L_{\rm plat} \approx 2\times10^{39}
	\frac{E_{48}^{5/6}R_{1000}^{2/3}}{M_{10}^{1/2}}
	\,{\rm erg\,s^{-1}},
\ee
with a duration of
\be
	t_{\rm plat} \approx 420\frac{M_{10}^{1/2}R_{1000}^{1/6}}{E_{48}^{1/4}}
	\,{\rm days},
\ee
which roughly matches \citet{LovegroveWoosley2013}. Comparison between $L_{\rm bo}$ and $L_{\rm plat}$ shows that breakout emission will be more conducive to detection by surveys.

An important constraint on observing the breakout from BH formation is that the shock must have sufficient energy to produce an observable signal. Estimating the breakout duration in the gas-dominated regime,
\be
	t_{\rm bo} = t_{\rm diff}
	= 7.8
	\frac{M_{10}^{0.20}R_{1000}^{2.19}}{\kappa_{0.34}^{0.62}E_{48}^{0.81}}
	\lp\frac{\rho_1}{\rho_*} \rp^{-0.25}{\rm days},
\ee
and the corresponding breakout luminosity is
\be
	L_{\rm bo} =  2.5\times10^{42}
	\frac{E_{48}^{5.22}}{\kappa_{0.34}^{1.12}M_{10}^{3.79}R_{1000}^{0.82}}
	\lp\frac{\rho_1}{\rho_*} \rp^{-0.95}{\rm erg\,s^{-1}}.
\ee
Setting this as $>10^{39}\,{\rm erg\,s^{-1}}$, the shock energy must obey
\be
	E> 2.2\times10^{47} 
	\kappa_{0.34}^{0.21}M_{10}^{0.73}R_{1000}^{0.16}
	\lp\frac{\rho_1}{\rho_*} \rp^{0.18}{\rm erg}.
	\label{eq:emin}
\ee
\citet{LovegroveWoosley2013} find values around this range, both above and below, and with generally more energy for a $15\,M_\odot$ ZAMS star in comparison to $25\,M_\odot$. This suggests that the breakout may not be detectable in all cases. An important focus for future simulations of collapsing stars is to better explore the full range of shock energies possible.

\section{Conclusions and Discussion}
\label{sec:conclusion}

I have investigated shock breakout during the collapse of a red supergiant producing a BH. The shock is generated by the hydrodynamic response of $\sim{\rm few}\times0.1M_\odot$ being carried away by neutrinos in the time before BH formation. This breakout flash has many distinctive characteristics that will help distinguish it from other potential short timescale transients \citep[e.g.,][]{Metzgeretal2009a,Metzgeretal2009b,Darbhaetal2010,Metzgeretal2010,Shenetal2010},
which I summarize as follows.
\begin{enumerate}
\item The breakout flash has a luminosity $L_{\rm bo}\approx10^{40}-10^{41}{\rm erg\,s^{-1}}$, as given by equation (\ref{eq:lbo}).
\item The flash duration is $t_{\rm bo}\approx3-10\,{\rm days}$ with $t_{\rm bo}\propto R_*^{2.16}$ because it is set by $t_{\rm diff}$ (rather than $t_{\rm lc}$ as in many SNe).
\item The breakout is in the thermal regime with an observed temperature of $T_{\rm obs}\approx10,000\,{\rm K}$.
\item The velocities should be relatively low with $v_{\rm max}\approx 200\,{\rm km\,s^{-1}}$, as given by equation (\ref{eq:vmax}).
\item The star likely retained most of its hydrogen envelope to have a sufficiently large radius for a bright and long lasting breakout, and this would be seen in the spectra.
\item The spectrum should be devoid of nucleosynthetic products from explosive burning.
\end{enumerate}
The source would peak in the ultraviolet with an absolute magnitude of roughly $-14.5$. The spectrum will be bright in blue and visual wavebands, making it well-suited for detection by wide-field, transient
surveys like the Palomar Transient Factory \citep[PTF;][]{Rauetal2009,Lawetal2009} and the Panoramic Survey Telescope and Rapid Response System \citep[Pan-STARRS;][]{Kaiseretal2002}. Future theoretical work should better quantify the time-dependent temperature and luminosity during this phase. If the progenitor is a blue supergiant or a Wolf-Rayet star rather than a red supergiant, then the smaller radius causes the shock breakout duration to be significantly shorter than what I summarize here. Nevertheless, the main properties can be estimated using Figure \ref{fig:luminosity}.

A critical issue I have explored, which typically does not occur for normal shock breakouts from SNe, is the impact of gas pressure and adiabatic expansion at low shock energies. It is shown that if the shock is sufficiently low energy, as quantified by equation (\ref{eq:emin}), then the breakout will not be observable. A critical question for future theoretical studies will therefore be to explore what exactly is the expected range of energies for this shock. This may depend on many factors, including the progenitor mass, neutron star equation of state, treatment of neutrino emission, and duration of neutrino emission from the protoneutron star prior to collapse to a BH. For example, \citet{LovegroveWoosley2013} find that a NS equation of state that favors a massive maximum mass ($\approx2.5\,M_\odot$) results in increased neutrino emission and a stronger shock. This shows how the detection and study of the signal described here could assist in addressing other fundamental questions in physics and astrophysics.

\acknowledgments
I thank Elizabeth Lovegrove and Stan Woosley for sharing their models and helpful discussions of their work. I also thank Christopher Kochanek, Evan O'Connor, and Christian Ott for feedback on previous drafts. This work was supported through NSF grants AST-1212170, PHY-1151197, and PHY-1068881, NASA ATP grant NNX11AC37G, and by the Sherman Fairchild Foundation.

\end{document}